\documentstyle[preprint,aps,epsf,floats,eqsecnum]{revtex}

\let\mdef\def
\mdef\etal{{\it et al.}}
\mdef\Tr{\mathop{\rm Tr}\nolimits}
\mdef\tr{\mathop{\rm tr}\nolimits}
\mdef\str{\mathop{\rm str}\nolimits}
\mdef\sdet{\mathop{\rm sdet}\nolimits}
\mdef\det{\mathop{\rm det}\nolimits}
\mdef\Str{\mathop{\rm Str}\nolimits}
\mdef\Sdet{\mathop{\rm Sdet}\nolimits}
\mdef\Det{\mathop{\rm Det}\nolimits}
\mdef\scr#1{{\cal #1}}
\mdef\eqb{begin{equation}}
\mdef\eqe{end{equation}}
\mdef\ecm{e\,{\rm cm}}
\mdef\MeV{{\rm \,MeV}}
\mdef\GeV{{\rm \,GeV}}
\mdef\mev{{\rm \,MeV}}
\mdef\gev{{\rm \,GeV}}
\mdef\qt{{\tilde q}}
\mdef\qbar{{\overline q}}
\mdef\qtbar{{\overline\qt}}
\mdef\phit{{\tilde\phi}}
\mdef\etat{{\tilde \eta}}
\mdef\mii{m_{ii}}
\mdef\mjj{m_{jj}}
\mdef\mij{m_{ij}}

\tighten

\begin{document}

\preprint{\vbox{\hbox{JHU--TIPAC--96016}\hbox{hep-ph/9610532}}}

\title{Quenched Chiral Perturbation Theory\\ for Vector Mesons}
\author{Michael Booth, George Chiladze and  Adam F.~Falk}
\address{Department of Physics and Astronomy,
The Johns Hopkins University\\
3400 North Charles Street, Baltimore, Maryland 21218 U.S.A.}

\date{October 1996}
\maketitle

\begin{abstract} 
We develop quenched chiral perturbation theory for vector mesons made
of light quarks, in the limit where the vector meson masses are much
larger than the pion mass.  We use this theory to extract the leading nonanalytic dependence of the vector meson masses on the masses of the light quarks.  By comparing with analogous quantities computed in ordinary
chiral perturbation theory, we estimate the size of
quenching effects, observing that in general they can be quite large.  This estimate is relevant to lattice simulations, where
the $\rho$ mass is often used to set the lattice spacing.
\end{abstract}


\section{Introduction}

Lattice QCD has made great progress in recent years, to the point
that the computation of the masses of the light hadrons with the accuracy of 
a few percent is within sight~\cite{Gottlieb:lat96}.  
Of course, such a precision requires that one understand well the various approximations, extrapolations and ans\"atze which underlie the calculation, in order to estimate, and then minimize, the error which they induce.  Without a doubt, the most significant of these, and by far the least well understood, are the corrections due to quenching, particularly as they affect the extrapolation of results to the physical values of the light quark masses.  

Quenching errors divide roughly into short and long distance parts.
At short distance, their primary effect is to change the running of the
coupling constant $\alpha_s$. This manifests itself, for instance, in
the wave functions of quarkonia, where it can
be studied in detail~\cite{El-Khadra:alphas1}.  The long distance effects are by their
nature more difficult to quantify.  One approach is to use quenched
chiral perturbation theory (QChPT), that is, the low energy effective theory of quenched QCD.  This idea was proposed by
Sharpe~\cite{Sharpe:QChPT1,Sharpe:QChPT2}, with the formalism
further developed by Bernard and
Golterman~\cite{BernGolt:QChPT1,BernGolt:QChPT2}.  The advantage of
using a version of chiral perturbation theory to study this problem is that
its predictions follow from the basic
properties of the underlying theory.  Moreover, ordinary chiral perturbation theory is known to
describe the low energy dynamics of unquenched QCD very well.

More recently, QChPT has been extended to describe the interactions of pions with baryons~\cite{Labrenz:QBary1,Labrenz:QBary2} and
heavy mesons~\cite{Booth:QChHQ,Sharpe:QHQ} in quenched QCD.  In this paper, we will
extend it further to describe the interactions of the light vector mesons, $\rho$, $K^*$ $\omega$ and $\phi$.  
Jenkins, Manohar and Wise~\cite{Manohar:VM1} have formulated ordinary chiral perturbation
theory for the vector mesons, treating the vector mesons as heavy
particles, in a manner similar to chiral perturbation theory for baryons~\cite{Jenkins:Bary}.  
We will adapt their formalism to quenched QCD.  Although we will adhere to most of their conventions, we will use a slightly modified Lagrangian which generalizes more easily to the quenched case and to an arbitrary number $N_f$ of light quarks.

What can we expect to learn from such an investigation?  The difficulty with low energy effective theories is that typically they contain a large number of undetermined coupling constants, all of which must be fixed from experiment.   In fact, only for that sector of chiral perturbation theory which describes the self-interactions of the pions are there enough data to carry out such a  phenomenological program beyond leading order.  For all other particles and interactions in the chiral Lagrangian, the data are scanty at best.  The situation in the quenched theory is even worse, since the only ``data'' available are extracted from lattice simulations.  Hence we are left in a situation where firm and accurate predictions are largely impossible to obtain.

Instead, we take a more modest approach.  First, we note that there are certain loop corrections which are predicted unambiguously by the lowest order Lagrangian, namely those that have a {\it nonanalytic\/} dependence on the light quark masses.  Such contributions to physical observables cannot be compensated by higher order terms in the chiral expansion.  Furthermore, this nonanalytic dependence is particularly sensitive to the long distance behavior of the theory, which is what we wish to investigate.  Second, we use the power counting rules which are implicit in chiral perturbation theory, by which all nonperturbative coupling constants are taken to be of order one (given certain other conventions).  With this assumption, we can ask whether the nonanalytic dependence of a physical quantity on the light quark masses is small or large.  If it is small, we enhance our confidence that the low energy theory is well behaved.  However, if it is large, then the applicability of the low energy theory to that quantity is called into question.  

The purpose of this investigation, then, is to explore whether QChPT is under reasonable control, by studying those quantities which the theory, at some level, predicts unambiguously.  We will find, in fact, that QChPT is quite badly behaved for the masses of the vector mesons, for apparently reasonable values of the nonperturbative coupling constants.  Whether this conclusion ought to be extended to quenched QCD itself is not completely clear, but at the very least, our results should sound a serious note of caution about quenched lattice simulations of the vector meson masses.

\section{Quenched chiral perturbation theory}

The extension of chiral perturbation theory to describe quenched QCD is by now standard.  Following Morel~\cite{Morel:QLogs}, 
for each quark $q^i$, a corresponding bosonic ``ghost'' quark $\tilde
q^i$ is introduced with identical mass, so that the fermion
determinant is canceled by the ghost determinant.  
These ghost quarks will then form mesons
with the true quarks and with themselves. As a result, the symmetry
group of the theory is enlarged from $SU(N_f)_L \times SU(N_f)_R$ to
the semi-direct product $(SU(N_f|N_f)_L \times
%SU(N_f|N_f)_R){\otimes}U(1)$.
SU(N_f|N_f)_R)\otimes U(1)$.

In this larger theory, the matrix of Goldstone bosons is
promoted to a supermatrix,
\begin{equation}
   \Pi = \left(\matrix{\pi&\chi^\dagger\cr
	  \chi&\tilde\pi\cr}\right),
\end{equation}where the quark/ghost content of the fields is
$\pi\sim q\qbar$, $\chi^\dagger\sim\qt\qbar$, $\chi\sim q\qtbar$ and
$\tilde\pi\sim\qt\qtbar$.  Each of these is actually an $N_f\times
N_f$ matrix; for example, for $N_f=3$, $\pi$ is the ordinary
pseudoscalar nonet
\begin{equation}
   \pi = 
   \pmatrix{\frac 1{\sqrt 2}\pi^0 + \frac 1{\sqrt 6}\eta&
   \pi^+ & K^+ \cr
   \pi^- & -\frac 1{\sqrt 2} \pi^0 + \frac 1{\sqrt 6}\eta&
   K^0 \cr
   K^- & {\overline{K}}^0 & -\sqrt{\frac 2 3} \eta\cr} 
   + \frac1{\sqrt 3}\, \eta'I_3\,.
\end{equation}
Note that $\chi$ and $\chi^\dagger$ are fermionic fields, while $\pi$
and $\tilde\pi$ are bosonic.

The Lagrangian of QChPT is given by 
\begin{eqnarray}\label{eqn:LBG}
   \scr{L}_{Q\chi} &=& {f^2\over 8} \left(
   \Str[\partial_\mu\Sigma\partial^\mu\Sigma^\dagger]
   +4B_0\,\Str[\scr{M}_{+}] \right)\nonumber\\
   &&\qquad+{1\over2}\left(A_0\Str[\partial_\mu\Pi]
   \Str[\partial^\mu\Pi]- M_0^2 \Str[\Pi]\Str[\Pi]\right).
\end{eqnarray}
Here $\Sigma = \xi^2$, $\xi = e^{i\Pi / f}$ 
(the normalization is such that $f_\pi\approx130\MeV$)
while
\begin{eqnarray}
   \scr{M} &=& \left(\matrix{M&0\cr 0&M\cr}\right), \\
   M &=& {\rm diag}(m_1,\ldots,m_{N_f})\,,
\end{eqnarray}
and $\scr{M}_\pm = \frac 12 (\xi^\dagger \scr{M}\xi^\dagger \pm \xi
\scr{M} \xi)$.  The ``supertrace'' $\Str$ is defined with a minus
sign for the ghost-antighost fields.  The chief difference between the
quenched and unquenched theories is the presence of the terms
involving $\Str[\Pi]=N_f^{1/2}\,(\eta'-\tilde\eta')$.  In the
unquenched theory they can be neglected because they describe the
dynamics of the $\eta'$ meson, which decouples from the theory.  But
quenching prevents the $\eta'$ from becoming heavy and decoupling, so
these terms must be retained in QChPT.  We have normalized $A_0$ and $M_0$ so that
they have no implicit dependence on $N_f$.

The propagators that are derived from this Lagrangian are the ordinary
ones, except for the flavor-neutral mesons, where the non-decoupling
of $\Pi_0$ leads to a curious double-pole structure.  It is convenient
to adopt a basis for the these mesons corresponding to $q_i\qbar_i$
and $\qt_i\qtbar_i$.  Then the propagator in the flavor-neutral sector
takes the form
\begin{equation}
  G_{ij}(p) = {\delta_{ij} \epsilon_i \over p^2 - \mii^2} 
  + {-A_0 p^2 + M_0^2 \over (p^2 - \mii^2)(p^2 - \mjj^2)}\,,
\end{equation}
where $\mii^2 = 2B_0 m_i$, and $\epsilon_i=1$ if $i$ corresponds to a
quark and $\epsilon_i=-1$ if $i$ corresponds to a ghost.  The second
term in the propagator is treated as a new vertex, the so-called
``hairpin,'' with the rule that it can be inserted only once on a
given line.  Note that this term can induce mixing between
quark-antiquark and ghost-antighost pairs.

The treatment for vector mesons is similar.  They are combined into
the supermatrix
\begin{equation}
  N_\mu = \left(\matrix{H_\mu & K_\mu \cr
	L_\mu &\widetilde H_\mu\cr}\right),
\end{equation}
with $H_\mu$ the usual matrix of vector mesons, which for $N_f=3$ is
\begin{equation}\label{eqn:nonet}
  H_\mu = \pmatrix{
  \frac 1{\sqrt 2}\rho_\mu^0+\frac 1{\sqrt6}\phi^8_\mu&
  \rho_\mu^+ & K_\mu^{*+} \cr
  \rho_\mu^- & -\frac 1{\sqrt 2} \rho_\mu^0 + \frac 1{\sqrt 6}\phi^8_\mu&
  K_\mu^{*0} \cr
  K_\mu^{*-} & {\overline{K}}_\mu^{*0} & -\sqrt{\frac 2 3} \phi_\mu^8\cr}
  +\frac1{\sqrt 3}\, S_\mu I_3\, .
\end{equation}
It is convenient to separate the neutral fields $\phi_\mu^8$ and
$S_\mu$ into $SU(3)$ octet and singlet pieces.  The actual mass
eigenstates, for $N_f=3$, are mixtures which correspond more closely
to the $\omega_\mu$ ($\bar u u+\bar d d$) and the $\phi_\mu$ ($\bar s
s$).

We will work in the ``heavy vector meson'' limit, previously used for
ordinary ChPT for vector mesons by Jenkins, Manohar and Wise~\cite{Manohar:VM1}.  
In this approximation, the vector mesons are treated as static fields, whose
velocity $v^\mu$ does not change when absorbing and emitting soft
pions.  The kinetic and mass terms for the vector meson fields are
given by
\begin{eqnarray}
  {\cal L}_{\rm kin} &=& - i \, \Str [ N_\mu^\dagger \left( v \cdot
 	{\cal D} \right) N^\mu ] 
  -i\,A_N  \Str [N_\mu^\dagger]
   \left( v \cdot{\cal D} \right) \Str[N^\mu]  \\
  {\cal L}_{\rm mass} &=& \mu_0\, \Str[N_\mu^\dagger] \Str[
  N^\mu] + \mu\, \Str[ N_\mu^\dagger N^\mu] +
 	\lambda_1\left(\Str[N_\mu^\dagger] \Str[ N_\nu 
  {\cal M}_+] + {\rm h.c.}\right) \\
	 &&\mbox{}+ \lambda_2 \, \Str[ \{N_\mu^\dagger, N^\mu\}{\cal M}_+ ]\,.
\end{eqnarray}
The covariant derivative is 
\begin{equation}
  {\cal D}_\mu N_\nu = \partial_\mu N_\nu + [V_\mu, N_\nu]\,, 
  \end{equation}
where
\begin{equation}
  V_\mu  = \frac 1 2
  \left(\xi \partial_\mu \xi^\dagger +\xi^\dagger \partial_\mu \xi \right)\,.
\end{equation}
There is also an axial combination,
\begin{equation}
  A_\mu =
  \frac i 2 \left(\xi \partial_\mu \xi^\dagger -\xi^\dagger \partial_\mu
  \xi \right)\,,
\end{equation}
which transforms homogeneously under the vector subgroup of the flavor
symmetry.

The propagator is somewhat different for the heavy vector mesons, but
there is still a double pole structure in the flavor-diagonal terms.
Let us neglect the $SU(3)$ violating mass 
%counter-terms 
terms 
proportional
to $\lambda_1$ and $\lambda_2$.  Then the propagator for the
off-diagonal mesons is
\begin{equation}
  \left(v^\mu v^\nu - g^{\mu\nu}\right){1\over v\cdot k -\mu}\,,
\end{equation}
where $k^\mu=P_N^\mu-(M_N-\mu)v^\mu$ is the ``residual momentum'' of
the vector meson $N$.  For the flavor-diagonal fields, the propagator
is
\begin{equation}
  G_{ij}^{\mu\nu}(k)=\left(v^\mu v^\nu - g^{\mu\nu}\right)
  \left[{\delta_{ij}\epsilon_i\over v\cdot k-\mu}
  +{-A_N v\cdot k+\mu_0\over(v\cdot k-\mu)^2}\right]\,.
\end{equation}
The vector mesons have a common ``residual mass'' $\mu$ at this order.
However, we could as easily choose to fix $\mu=0$, absorbing it all
into the static phase associated with the heavy vector meson
propagator.  We will do this from here on.

There are four invariant interactions between the vector mesons and
the Goldstone bosons:
\begin{eqnarray}
{\cal L}_{\rm int} &=& 
  ig_1 \Str[N_\mu^\dagger]\Str[N_\nu A_\lambda]\,
	v_\sigma \epsilon^{\mu\nu\lambda\sigma} + {\rm h.c.} \nonumber\\
  &&\mbox{}+
  ig_2 \Str[\{N_\mu^\dagger,N_\nu\}A_\lambda]\,
   v_\sigma\epsilon^{\mu\nu\lambda\sigma} \nonumber\\
  && \mbox{}+
  ig_3 \Str[N_\mu^\dagger]\Str[N_\nu]\Str[A_\lambda]\, 
	v_\sigma\epsilon^{\mu\nu\lambda\sigma} \nonumber\\
  && \mbox{}+
  ig_4 \Str[N_\mu^\dagger N_\nu]\Str[A_\lambda] \,
	v_\sigma\epsilon^{\mu\nu\lambda\sigma}\,.
\end{eqnarray}
Note that we could ``unquench'' the theory simply by removing the
ghost fields and the flavor-singlet $\eta'$, which then would receive
a large mass from the anomaly.  Only the first two interaction terms
in ${\cal L}_{\rm int}$ would remain, corresponding to the two
interactions considered in Ref.~\cite{Manohar:VM1}.

\section{Nonanalytic corrections to vector meson masses}

The interactions in the chiral Lagrangian effect the masses of the
vector mesons in two ways.  First, there are explicit $SU(3)$
violating terms which correspond to local interactions in the
theory.  One cannot say much about these terms, except to use the
available symmetries to constrain them at any given order in the
chiral expansion.  Second, there are ``long distance'' corrections
which arise from the infrared parts of loop integrals computed using
the Lagrangian at leading order.  These corrections have a nonanalytic
dependence on the light quark masses $m_i$, usually of the form
$\sqrt{m_i}$ or $m_i\ln m_i$, and thus cannot be canceled by
counterterms at any order in the expansion.  It is useful to study
these corrections, because, while they may not be dominant in general, they at
least set a scale for $SU(3)$ violation.  As discussed earlier, such
terms also correspond to the part of the theory where the effects of
quenching are the least understood.  Moreover, they may be dominant
for sufficiently small quark masses.

We will use the interaction Lagrangian ${\cal L}_{\rm int}$ to compute
the nonanalytic corrections to the vector meson masses for
$N_f=1,2,3$.  We will then compare our result with analogous
corrections computed in the unquenched theory.  In each case, we will
truncate the expansion to include nonanalytic terms of order $m_q^{1/2}$, $m_q\ln m_q$ and $m_q^{3/2}$, but not those of order $m_q^2\ln m_q$.  Where we
do not neglect them, we treat the dimensionful parameters
$\mu_0$ and $M_0^2$ as being formally of order $m_q$.

\subsection{$N_f=1$}

\begin{figure}
\epsfxsize=12cm
\hfil\epsfbox{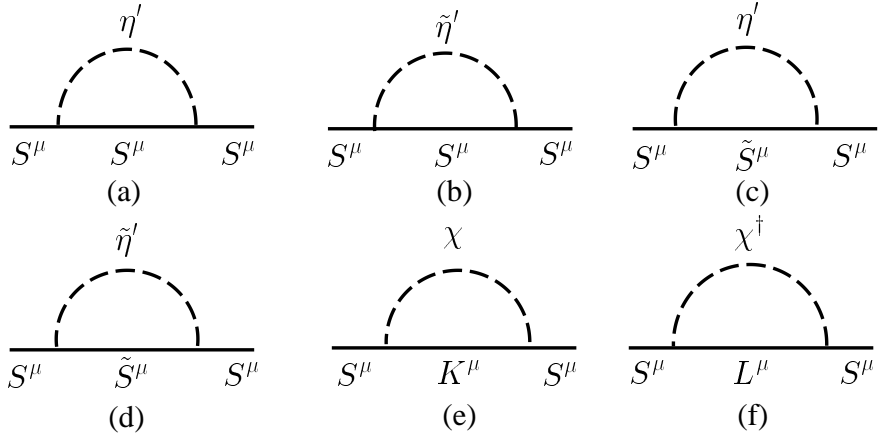}\hfill
\caption{Feynman diagrams with no hairpin insertions.}
\label{fig:0hair}
\end{figure}

The case $N_f=1$ is particularly simple, since there are only
flavor-singlet mesons.  Since it is the flavor-singlet sector which is
the source of most complications in QChPT, we work out this case in
detail as a pedagogical example.

We begin with the propagator for the Goldstone fields.  For the
$\eta'$ and $\tilde\eta'$, this is given in matrix form by
\begin{equation}
  {1\over p^2-m^2_{\eta'}}\pmatrix{1&0\cr 0&-1}
  +{-A_0 p^2+M_0^2\over(p^2-m^2_{\eta'})^2}\pmatrix{1&1\cr 1&1}\,.
\end{equation}
Note the relative minus sign in the non-hairpin parts of the $\eta'$
and $\tilde\eta'$ propagators.  The off-diagonal terms mix the $\eta'$
and the $\tilde\eta'$.  The propagator for the fermionic $\chi$ field
is simply
\begin{equation}
  {1\over p^2-m^2_{\eta'}}\,.
\end{equation}
The vector meson propagators are similar.  Choosing $\mu=0$, in the
bosonic sector we find
\begin{equation}
  \left(v^\mu v^\nu-g^{\mu\nu}\right){1\over v\cdot k}\pmatrix{1&0\cr 0&-1}
  +\left(v^\mu v^\nu-g^{\mu\nu}\right){-A_N v\cdot k+\mu_0\over
  (v\cdot k)^2}\pmatrix{1&1\cr 1&1}\,.
\end{equation}
The propagators for the fermions $L_\mu$ and $K_\mu$ are
\begin{equation}
   \left(v^\mu v^\nu-g^{\mu\nu}\right){1\over v\cdot k}\qquad
   {\rm and}\qquad
   -\left(v^\mu v^\nu-g^{\mu\nu}\right){1\over v\cdot k}\,,
\end{equation}
respectively.  A careful study of the ordering of creation and
annihilation operators in time-ordered products shows that fermion
loops containing an $L_\mu$ and a $\chi^\dagger$ come with the usual
minus sign, but those with a $K_\mu$ and a $\chi$ do not.  In the
Feynman diagrams which we will consider, this effectively cancels the
extra minus sign in the $K^\mu$ propagator.

Next, we expand the interaction Lagrangian ${\cal L}_{\rm int}$.
Since we will be computing the corrections to the real vector meson
self-energy at one loop, we only keep terms with a least one factor of
$S_\mu$.  Each of the six relevant couplings has a coefficient which
is a linear combination of the $g_i$'s.  Suppressing the common factor
of $v_\sigma\epsilon^{\mu\nu\lambda\sigma}$, we find the coefficients
\begin{eqnarray}
   S_\mu^\dagger S_\nu\partial_\lambda\eta'\quad &:&
     \qquad 2g_1+2g_2+g_3+g_4\nonumber\\
   S_\mu^\dagger S_\nu\partial_\lambda\tilde\eta'\quad &:&
     \qquad -g_3-g_4\nonumber\\
   S_\mu^\dagger \tilde S_\nu\partial_\lambda\eta'\quad &:&
     \qquad -g_1-g_3\nonumber\\
   S_\mu^\dagger \tilde S_\nu\partial_\lambda\tilde\eta'\quad &:&
     \qquad -g_1+g_3\nonumber\\
   S_\mu^\dagger K_\nu\partial_\lambda\chi+{\rm h.c.}\quad &:&
     \qquad g_1+g_2\nonumber\\
   S_\mu^\dagger L_\nu\partial_\lambda\chi^\dagger+{\rm h.c.}\quad &:&
     \qquad -g_1-g_2\,.
\end{eqnarray}

\begin{figure}
\epsfxsize=12cm
\hfil\epsfbox{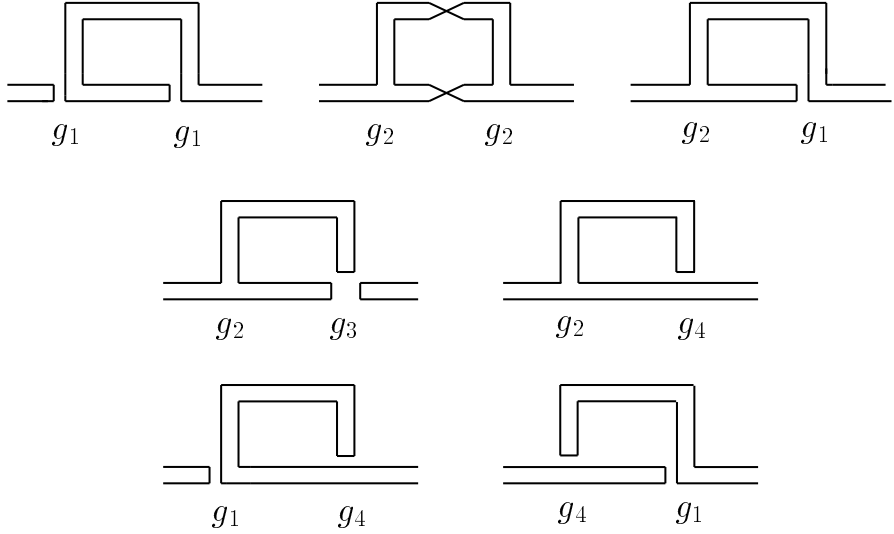}\hfill
\caption{Quark flow diagrams which contribute to $\delta M_S$ in the
quenched approximation.} 
\label{fig:flow0h}
\end{figure}

There are four types of graphs which contribute to the $S_\mu$
self-energy: those with no hairpin insertions, those with a single
hairpin insertion (on either the vector meson or the Goldstone boson
line), and those with two hairpin insertions (one on each internal
line).  The six Feynman diagrams with only ``ordinary'' propagators
are shown in Fig.~\ref{fig:0hair}.  The same loop integral appears in
all the graphs,
\begin{equation}
  \delta M_S\propto I_1(m_{\eta'})=-{1\over12\pi f^2}m_{\eta'}^3
 +({\rm analytic\ in\ }m^2_{\eta'}) \,.
\end{equation}
The six diagrams contribute to $\delta M_S$ with the following
coefficients of $I_1(m_{\eta'})$:
\begin{eqnarray}
   \ref{fig:0hair}({\rm a})\quad &:& 
      \qquad (2g_1+2g_2+g_3+g_4)^2\nonumber\\
   \ref{fig:0hair}({\rm b})\quad &:& \qquad -(-g_3-g_4)^2
      \nonumber\\
   \ref{fig:0hair}({\rm c})\quad &:& \qquad -(-g_1-g_3)^2
      \nonumber\\
   \ref{fig:0hair}({\rm d})\quad &:& \qquad (-g_1+g_3)^2
      \nonumber\\
   \ref{fig:0hair}({\rm e})\quad &:& \qquad -(g_1+g_2)^2
      \nonumber\\
   \ref{fig:0hair}({\rm f})\quad &:& \qquad -(-g_1-g_2)^2\,.
\end{eqnarray}
The minus signs in graphs \ref{fig:0hair}(b) and \ref{fig:0hair}(c) are from the $\eta'$ and
$\tilde S_\mu$ propagators; the minus signs in \ref{fig:0hair}(e) and \ref{fig:0hair}(f)
are from fermion loops.  Summing them up, we find
\begin{equation}\label{eq:ms1}
   \delta M_S = (2g_1^2+2g_2^2+4g_1g_2+4g_1g_4+4g_2g_3+4g_2g_4)\,
   I_1(m_{\eta'})+\dots
\end{equation}
Before we compute the remaining graphs, it is interesting to see which
combinations of the $g_i$'s contribute to $\delta M_S$.  Since the role
of the ghost quarks is to implement quenching by canceling the
contributions of closed quark loops, the terms which remain after the
diagrams have been added should be just those for which no quark loop is
necessary.  In Fig.~\ref{fig:flow0h} we present ``quark flow'' diagrams
corresponding to the combinations of coupling constants which
contribute.  It is an easy exercise to check that for the combinations
which do not appear in
$\delta M_S$, any quark flow diagram must contain a closed quark loop.

We pause for a moment to consider the dependence of the various terms
on the number of colors $N_c$.  In the large $N_c$ limit, the coupling
constants scale with $N_c$ as follows:
\begin{equation}
  g_1\sim{1\over N_c}\,,\quad g_2\sim 1\,,\quad
  g_3\sim{1\over N_c^2}\,,\quad g_4\sim{1\over N_c}\,.
\end{equation}
Hence, it would be tempting to argue that the term proportional to
$g_2^2$ in Eq.~(\ref{eq:ms1}) is dominant in this limit.  However, we
note from Fig.~\ref{fig:flow0h} that the corresponding quark (and
color) flow diagram is not planar, so it is also suppressed by
$1/N_c$.  Of course, this is what we would expect, since in the
$N_c\to\infty$ limit the spectrum of the field theory is composed of
noninteracting mesons, so Goldstone boson loops should have no effect
on the vector meson mass.  The lesson is that we must consider more
than the coupling constants when extracting the large $N_c$ dependence
of a term in $\delta M$.

\begin{figure}
\epsfxsize=10cm
\hfil\epsfbox{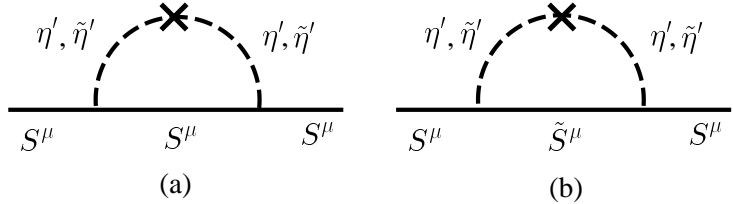}\hfill
\caption{Feynman diagrams with a hairpin insertion on the $\eta'$
propagator.}
\label{fig:1haireta}
\end{figure}

The next class of graphs are those with a hairpin insertion on the
$\eta'$ line, as shown in Fig.~\ref{fig:1haireta}.
In such
a graph, there can be two $\eta'$ propagators, two $\tilde\eta'$
propagators, or (in two ways), an $\eta'$ and an $\tilde\eta'$.  So from
the graph in Fig.~\ref{fig:1haireta}(a), we find a result which is
proportional to
\begin{equation}
  (2g_1+2g_2+g_3+g_4)^2 + 2(2g_1+2g_2+g_3+g_4)(-g_3-g_4)
  +(-g_3-g_4)^2=(2g_1+2g_2)^2\,,
\end{equation}
and from Fig.~\ref{fig:1haireta}(b),
\begin{equation}
  -\left[(-g_1-g_3)^2 + 2(-g_1-g_3)(-g_1+g_3)+(-g_1+g_3)^2\right]
  =-(-2g_1)^2\,,
\end{equation}
where the minus sign in the second expression comes from the $\tilde
S_\mu$ propagator.  Adding these and including the loop integral, we
obtain
\begin{equation}
  \delta M_S = \dots + (4g_2^2+8g_1g_2)I_2(m_{\eta'})+\dots\,,
\end{equation}
where
\begin{equation}
  I_2(m_{\eta'})=-{1\over12\pi f^2}\left({3\over2}M_0 m_{\eta'} 
  -{5\over2}A_0 m_{\eta'}^3\right)
  +({\rm analytic\ in\ }m^2_{\eta'})\,.
\end{equation}

\begin{figure}
\epsfxsize=10cm
\hfil\epsfbox{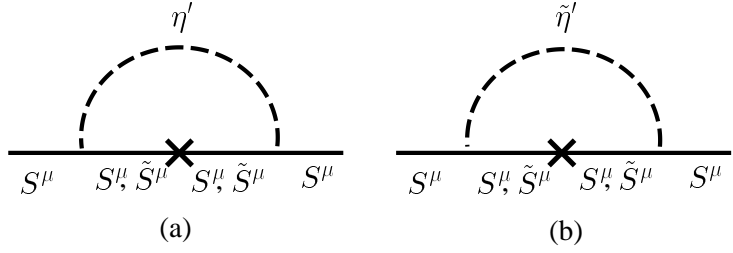}\hfill
\caption{Feynman diagrams with a hairpin insertion on the vector meson
propagator.}
\label{fig:1hairvec}
\end{figure}

Similarly, we can have a single hairpin insertion on the vector meson
line, as shown in Fig.~\ref{fig:1hairvec}.
Here the graph in
Fig.~\ref{fig:1hairvec}(a) is proportional to
\begin{equation}
  (2g_1+2g_2+g_3+g_4)^2 + 2(2g_1+2g_2+g_3+g_4)(-g_1-g_3)
  +(-g_1-g_3)^2=(g_1+2g_2+g_4)^2\,,
\end{equation}
when the $S_\mu$ and $\tilde S_\mu$ propagators are included.  The graph
in Fig.~\ref{fig:1hairvec}(b) is proportional to
\begin{equation}
  -\left[(-g_3-g_4)^2 + 2(-g_3-g_4)(-g_1+g_3)+(-g_1+g_3)^2\right]
  =-(-g_1-g_4)^2\,,
\end{equation}
where the minus sign is from the $\tilde\eta'$ propagator.  Combining
these terms, we find
\begin{equation}
  \delta M_S = \dots + (4g_2^2+4g_1g_2+4g_2g_4)I_3(m_{\eta'})+\dots\,,
\end{equation}
where
\begin{equation}
  I_3(m_{\eta'})={1\over12\pi f^2}\left({3\over2\pi}\mu_0 m^2_{\eta'}
  \ln m_{\eta'}^2 +A_N m_{\eta'}^3\right)
  +({\rm analytic\ in\ }m^2_{\eta'})\,.
\end{equation}

\begin{figure}
\epsfxsize=5cm
\hfil\epsfbox{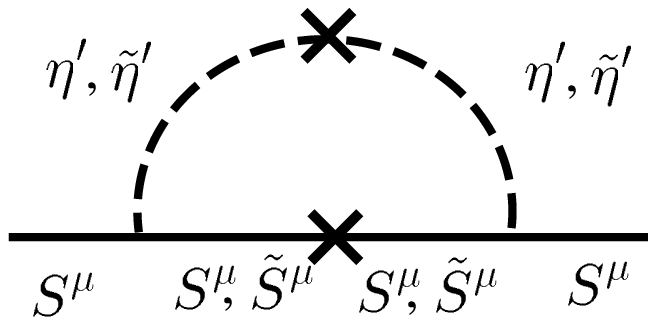}\hfill
\caption{Feynman diagrams with two hairpin insertions.}
\label{fig:2hair}
\end{figure}

Finally, there are graphs with two hairpin insertions, one each on the
$\eta'$ and $S_\mu$ propagators.  The sixteen possible graphs are
illustrated compactly in Fig.~\ref{fig:2hair}.
It is straightforward to enumerate them as before, and we find
\begin{equation}
  \delta M_S = \dots + 4g_2^2\,I_4(m_{\eta'})\,,
\end{equation}
where
\begin{eqnarray}
  I_4(m_{\eta'})&=&{1\over12\pi f^2}\left[
  {3\over2\pi}\mu_0 M^2_0\left(1+\ln m_{\eta'}^2\right)
  -{3\over2\pi}\mu_0 A_0 m_{\eta'}^2\left(1+2\ln m_{\eta'}^2\right)
  \right.\nonumber\\
  &&\qquad\qquad
  \left.\mbox{}+{3\over2}A_N M_0^2 m_{\eta'}-{5\over2}A_NA_0 m_{\eta'}^3
  \right]+({\rm analytic\ in\ }m^2_{\eta'})\,.
\end{eqnarray}
In Fig.~\ref{fig:hairpinflow} we present quark flow diagrams
corresponding to the surviving diagrams with hairpin insertions, in
which there is no closed quark loop.  Again, one can convince oneself
that all other combinations of coupling constants require quark loops,
and hence do not contribute in the quenched theory.

\subsection{$N_f>1$}

\begin{figure}
\epsfxsize=12cm
\hfil\epsfbox{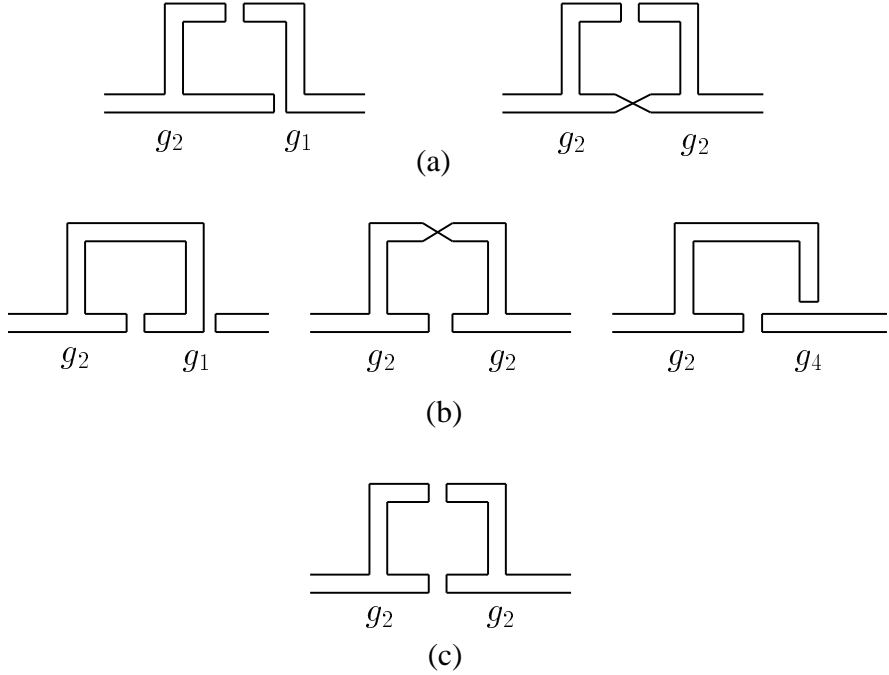}\hfill
\caption{Quark flow diagrams corresponding to hairpin insertions.  
(a) hairpin insertion on the $\eta'$ line; 
(b) hairpin insertion on the $S^\mu$ line; (c) two hairpin insertions.}
\label{fig:hairpinflow}
\end{figure}

If there is more than one light flavor, then the situation is
complicated by the fact that there is more than one flavor of vector
meson.  However, we will see the result in this case can be obtained
largely by considering the quark flow diagrams in Figs.~\ref{fig:flow0h} and
\ref{fig:hairpinflow} for a single flavor.

The matrix of vector mesons $H_\mu$ has $N_f^2$ entries, most
conveniently enumerated in the basis $q_i\qbar_j$.  In this basis,
there are three types of contributions to the mass matrix: $\delta
M_{ij}$, $i\ne j$, the correction to the masses of the off-diagonal meson
$q_i\qbar_j$; $\delta M_{ii}$, the correction to the mass of the
flavor-diagonal meson $q_i\qbar_j$; and $\delta M_{ii-jj}$, $i\ne j$,
the correction which mixes the flavor-diagonal mesons $q_i\qbar_i$ and
$q_j\qbar_j$.  Each of these corrections arises from quark flow
diagrams with a particular structure.

The quark flow diagrams in Figs.~\ref{fig:flow0h} and
\ref{fig:hairpinflow} can be divided into two classes: those in which
the quark flavor lines run through from the incoming vector meson to
the outgoing one, and those in which they do not.  In the first class
are the $g_1 g_2$ and $g_2 g_4$ terms in Fig.~\ref{fig:flow0h} and the
$g_2^2$ terms in Figs.~\ref{fig:hairpinflow}(a) and
\ref{fig:hairpinflow}(b).  The others are in the second class.
Because flavor is conserved in those diagrams in which the quark lines
flow through, they contribute to $\delta M_{ij}$ and $\delta M_{ii}$,
but not to $\delta M_{ii-jj}$.  In the rest of the diagrams, the quark
lines in the incoming meson must annihilate in the graph, so they
cannot contribute to the off-diagonal $\delta M_{ij}$; however, they
do contribute to $\delta M_{ii}$ and $\delta M_{ii-jj}$.

The only other differences in the $N_f>1$ case are the somewhat more complicated integrals which arise when a double-pole propagator with $\mii\ne\mjj$ appears in a diagram with a hairpin insertion on the Goldstone boson line.  The solution is written in terms of integrals with the
following nonanalytic dependence on $x^2$ and $y^2$:
\begin{eqnarray}
  I_1(x) &=& -{1\over12\pi f^2}\,x^3\nonumber\\
  I_2(x,y) &=& {1\over x^2-y^2}\left[(M_0^2-A_0 x^2)I_1(x)
      -(M_0^2-A_0 y^2)\,I_1(y)\right]\\
  I_3(x) &=& {1\over12\pi f^2}\,
      \left[{3\over2\pi}\mu_0 x^2\ln x^2 + A_Nx^3\right]\nonumber\\
  I_4(x,y) &=& {1\over x^2-y^2}\left[(M_0^2-A_0 x^2)I_3(x)
      -(M_0^2-A_0 y^2)\,I_3(y)\right]\,.
\end{eqnarray}
The integrals $I_2$ and $I_4$ have the degenerate limits
\begin{eqnarray}
  I_2(x,x)&=&-{1\over12\pi f^2}\,\left[
  {3\over2} M^2_0 x-{5\over2}A_0 x^3\right]\nonumber\\
  I_4(x,x)&=&{1\over12\pi f^2}\,\left[
  {3\over2\pi}\mu_0 M^2_0\left(1+\ln x^2\right)
  -{3\over2\pi}\mu_0 A_0 x^2\left(1+2\ln x^2\right)\right.\nonumber\\
  &&\qquad\qquad\qquad
  \left.\mbox{}+{3\over2}A_N M_0^2 x-{5\over2}A_NA_0 x^3\right]\,.
\end{eqnarray}
The integrals $I_i$ will depend on the variables $\mij=[B_0(m_i+m_j)]^{1/2}$.

For the masses of the off-diagonal vector mesons, we find the correction
\begin{eqnarray}
  \delta M_{ij} &=& 4g_1g_2\,I_1(\mij)+2g_2g_4\,
  \left[I_1(\mii)+I_1(\mjj)\right]\nonumber\\
  &&\quad\mbox{}
  +g_2^2\,\left[I_2(\mii,\mii)+2I_2(\mii,\mjj)+I_2(\mjj,\mjj)\right]
  +4g_2^2\,I_3(\mij)\,.
\end{eqnarray}
The flavor-conserving diagonal mass matrix is perturbed by
\begin{eqnarray}
  \delta M_{ii} &=& \left( 2g^2_1+2g_2^2+4g_1g_2
  +4g_1g_4+4g_2g_3+4g_2g_4\right)I_1(\mii)
  \nonumber\\
  &&\quad\mbox{}+\left(4g_2^2+8g_1g_2\right)I_2(\mii,\mii)
  +\left(4g_2^2+4g_1g_2+4g_2g_4\right)I_3(\mii)\nonumber\\
  &&\quad\mbox{}+4g_2^2\,I_4(\mii,\mii)\,.
\end{eqnarray}
Finally, the contribution to mixing between flavor-diagonal mesons is
\begin{eqnarray}
  \delta M_{ii-jj} &=& \left(g_1^2+2g_2g_3
  +2g_1g_4\right)\,\left[I_1(\mii)+I_1(\mjj)\right]
  +2g_2^2\,I_1(\mij)\nonumber\\
  &&\quad\mbox{}+2g_1g_2\left[I_2(\mii,\mii)
  +2I_2(\mii,\mjj)+I_2(\mjj,\mjj)\right]\nonumber\\
  &&\quad\mbox{}+\left(2g_1g_2+2g_2g_4
  \right)\,\left[I_3(\mii)+I_3(\mjj)\right]
  +4g_2^2\,I_4(\mii,\mjj)\,.
\end{eqnarray}

\subsection{Vector meson masses in ordinary chiral perturbation theory}

We would like to compare these results with the nonanalytic
corrections to the vector meson masses computed using ordinary chiral perturbation theory.
The unquenched case, as discussed in Ref.~\cite{Manohar:VM1}, is somewhat
simpler.  There are only the fields $H^\mu$ and $\pi$, with no $\eta'$
field and no hairpin insertions.  The interaction
Lagrangian is simply
\begin{eqnarray}
  {\cal L}_{\rm \chi} &=& ig_1\Tr[H_\mu^\dagger]
  \Tr[H_\nu A_\lambda]\,
  v_\sigma \epsilon^{\mu\nu\lambda\sigma} + {\rm h.c.} \nonumber\\
  &&\mbox{}+ig_2\Tr[\{N_\mu^\dagger,N_\nu\}A_\lambda]\,
  v_\sigma\epsilon^{\mu\nu\lambda\sigma}\,.
\end{eqnarray}
Our normalization conventions are chosen to correspond as closely as
possible to the quenched Lagrangian, and they differ slightly from
those of Ref.~\cite{Manohar:VM1}.  A straightforward calculation yields the
nonanalytic corrections to the vector meson mass matrix for
$N_f=1,2,3$.  For $N_f=1$, there is only the singlet $S$, and
\begin{equation}
  \delta M_S=0\,.
\end{equation}
For $N_f=2$, there are the isotriplet $\rho$ and the singlet $S$.
Ignoring isospin violation, we find
\begin{eqnarray}
  \delta M_\rho &=& 2\left(g_1+g_2\right)^2 I_1(m_\pi)\nonumber\\
  \delta M_S &=& 6\left(g_1+g_2\right)^2 I_1(m_\pi)\nonumber\\
  \delta M_{\rho-S} &=& 0\,.
\end{eqnarray}
For $N_f=3$, we have
\begin{eqnarray}
  \delta M_\rho &=& 3\left(g_1+{2\over3}g_2\right)^2 I_1(m_\pi)
    +g_2^2\left[{2\over3}I_1(m_\pi)+2I_1(m_K)
    +{2\over3}I_1(m_\eta)\right]\nonumber\\
  \delta M_{K^*} &=& 3\left(g_1+{2\over3}g_2\right)^2 I_1(m_K)
    +g_2^2\left[{3\over2}I_1(m_\pi)+{5\over3}I_1(m_K)
    +{1\over6}I_1(m_\eta)\right]\nonumber\\
  \delta M_{\phi^8} &=& 3\left(g_1+{2\over3}g_2\right)^2 I_1(m_\eta)
    +g_2^2\left[2I_1(m_\pi)+{2\over3}I_1(m_K)
    +{2\over3}I_1(m_\eta)\right]\nonumber\\
  \delta M_S &=& 3\left(g_1+{2\over3}g_2\right)^2 
    \left[3I_1(m_\pi)+4I_1(m_K)+I_1(m_\eta)\right]\nonumber\\
  \delta M_{\phi^8-S} &=& \sqrt3 g_2\left(g_1+{2\over3}g_2\right)
    \left[3I_1(m_\pi)-2I_1(m_K)-I_1(m_\eta)\right]\,.
\end{eqnarray}
These expressions for $N_f=3$ correspond to those of Ref.~\cite{Manohar:VM1} if we make the replacement
\begin{equation}
  g_1\to{1\over\sqrt3}\left(g_1-{2\over\sqrt3}g_2\right)\,.
\end{equation}

\section{Phenomenology}
\label{sec:phenom}

The coefficients which appear in the chiral Lagrangians, both quenched
and unquenched, are nonperturbative parameters which are difficult to calculate from first principles.  However, we
need to have some estimate of their size if we are to use our
expressions for phenomenology.

The parameters $A_0$ and $M_0^2$ describing the $\eta'$
self-interaction may be studied on the lattice through the hairpin
vertex, the anomalous scaling of $m_\pi$, and the topological
susceptibility $\chi_t$ of the pure gauge theory.  The effect of the
hairpin on $m_\pi$ is parameterized by a quantity $\delta$, which is
extracted both indirectly and directly.  The values of $\delta$
measured on the lattice vary widely~\cite{Sharpe}, in the approximate range
$0<\delta<0.3$.  With our normalization of the terms in the effective
Lagrangian, the relationship between $M_0$ and $\delta$ is
\begin{equation}\label{eqn:delt}
  M_0^2=8\pi f^2\delta\,,
\end{equation}
from which we find $0<M_0<350\mev$.  

On the other hand, $M_0$ is also related to the topological
susceptibility $\chi_t$ via the Witten-Veneziano formula~\cite{Witten,Venez},
\begin{equation}\label{eqn:top}
  M_0^2={4\over f^2}\,\chi_t\,.
\end{equation}
One must be careful in taking this formula from the original
literature, since $\chi_t$ is evaluated in the limit $N_c\to\infty$.
In this limit $A_0=0$, that is, the wavefunction renormalization due to the hairpin vanishes.  Hence there is no distinction between writing the left-hand side of Eq.~(\ref{eqn:top}) as $M_0^2$ or, instead, as $M_0^2/(1+N_fA_0)$.  If we were to take this latter expression, then it would appear that
together Eqs.~(\ref{eqn:delt}) and (\ref{eqn:top}) determine {\it
both\/} $M_0$ and $A_0$.\footnote{In the notation of Duncan~\etal~\cite{Duncan},
the Lagrangian is written in terms of the parameters $A=1+N_fA_0$ and
$m_0^2=M_0^2N_f/(1+N_fA_0)$.  They take $M_0^2/(1+N_fA_0)$ in Eq.~(\ref{eqn:top}), which corresponds to writing it as $m_0^2=4N_f\chi_t/f^2$ instead of as
$Am_0^2=4N_f\chi_t/f^2$.  We agree that Eq.~(\ref{eqn:delt})
corresponds to their expression $Am_0^2=8N_f\pi^2f^2\delta$.}  However, it is inconsistent to keep
the wavefunction renormalization due to $A_0$ in one of these
relations but not in the other.  (If we do so, we are led to the contradictory conclusion that $A_0\sim1/N_f$, while really $A_0$ is independent of $N_f$.)  Instead, Eqs.~(\ref{eqn:delt}) and
(\ref{eqn:top}) yield two independent determinations of the single
variable $M_0$.  Unfortunately, these determinations are not
consistent with each other.  Lattice calculations give
$\chi_t\approx(180\mev)^4$~\cite{Duncan}, from which we find $M_0\approx500\mev$.
As a compromise, we will take $M_0=400\mev$ in our estimates below.

We have less information on the parameter $A_0$.  Theoretical
prejudice and the large $N_c$ limit would lead one to believe that
$A_0$ is small, but perhaps the inconsistency in the determination of
$M_0$ is a hint that this limit does not work well here.  In his
recent review~\cite{Sharpe}, Sharpe fits values for a parameter $\alpha_\Phi=A_0/3$ from a number of lattice groups, with the preferred value $A_0\sim0.2$. This is the value we will use in the numerical estimates below,
keeping in mind that it is probably highly uncertain.

We will also assume that $f\approx130\mev$ and $B_0\approx4\pi f$ are
the same in quenched and unquenched chiral perturbation theory, and
the same for the coupling constants $g_1$ and $g_2$.  All of the
couplings $g_i$ are expected to be of order one, with unknown sign.  We
will assume that there are no dramatic accidental cancellations
between different terms, and will take linear combinations of the
$g_i$'s to be of order one as well.  Finally, we will neglect the vector
meson hairpin, setting $A_N=0$ and $\mu_0=0$.  
Our primary motivation for this is simplicity;\footnote{
We note that in the real world, the size of 
$m_\rho - m_\omega$ suggests that these terms are indeed small.} 
complete expressions in which these terms are included are given above.

\subsection{The quenched $\rho$ mass}

We now use these results to extract information about the effect of
quenching in the determination of the $\rho$ mass on the lattice.
Quenched chiral perturbation theory gives us information about the
nonanalytic dependence of $m_\rho$ on the light quark masses.  Let
us consider a charged $\rho$ in a theory with $N_f=2$, the simplest case
in which there is no annihilation channel.  We will take the two quarks
to have equal mass $m_q$.  Then the expansion of $m_\rho$ in terms of
$m_q$ takes the form
\begin{equation}
  m_\rho=\mu + C_{1/2}\,m_\pi + C_1\, m_\pi^2 + C_{3/2}\, m_\pi^3+\dots\,,
\end{equation}
where the coefficients $C_i$ are given by
\begin{eqnarray}
  C_{1/2}&=&-{M_0^2g_2^2\over2\pi f^2}\sim -1.5 \nonumber\\
  C_1&=&{\lambda_2\over B_0}\sim0.1\times(100\mev)^{-1}\nonumber\\
  C_{3/2}&=&-{1\over12\pi f^2}\left[4g_2(g_1+g_4)-
      10\,g_2^2 A_0\right]\sim-0.1\times(100\mev)^{-2}\,,
\end{eqnarray}
and $m_\pi$ is related to $m_q$ by $m_\pi^2=2B_0m_q$. For comparison, the leading non-analytic term in the unquenched case is $C^\chi_{3/2}\,m_\pi^3$, with
\begin{equation}
  C^\chi_{3/2} = -{2(g_1+g_2)^2\over12\pi f^2}\sim -0.1\times(100\mev)^{-2}\,.
\end{equation}

\begin{figure}
\epsfxsize=12cm
\hfil\epsfbox{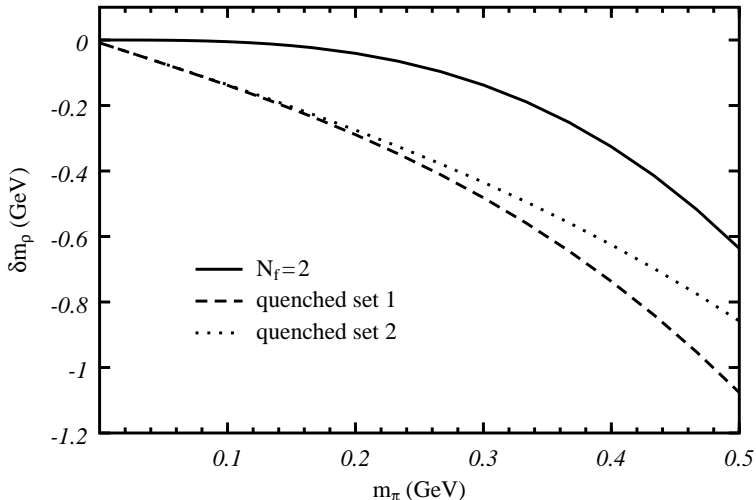}\hfill
\caption{Correction $\delta m_\rho$ to the $\rho$ mass, as a function of $m_\pi$, for $N_f=2$.  The solid line is the result for an unquenched theory; the dashed
and dotted lines are for quenched theories with $M_0=0.4\GeV$, $A_0=0$ and
$M_0=0.1\GeV$, $A_0=0.2$, respectively.  We have taken all $g_i=0.75$.}
\label{fig:mrho}
\end{figure}

Of course, the numerical estimates are intended only to give a sense
of the order of magnitude of the terms.  The $C_{1/2}m_\pi$ term is
large, and its sign is predicted by the quenched theory.  Moreover,
it is the leading term for small quark mass.  The
other corrections are at the 10\% level.  A fit to lattice data would
determine the $C_i$'s and check whether they are of the expected size.
In Fig.~\ref{fig:mrho} we present the corrections
to $m_\rho$ as a function of $m_\pi$, for a typical set of coupling constants $g_i$ of order one.  We show results for two
different choices of the couplings $M_0$ and $A_0$.  We also show the unquenched result.  As 
Fig.~\ref{fig:mrho} illustrates, it is quite generic to find extremely substantial nonanalytic corrections to $m_\rho$ for $m_\pi\agt400\mev$.  Of course, the possibility of such large quenching corrections in chiral perturbation theory does little to enhance our confidence in the precise extraction of $m_\rho$ from lattice simulations of quenched QCD. 

It is also useful to consider the case of unequal quark masses,
and to study the quenched chiral corrections to the mass splittings.
An interesting quantity to study is
$(m_{K^*}-m_\rho)/(m_K^2-m_\pi^2)$, which is constant in the chiral
limit at leading order.  The expression simplifies if we set $m_\pi=0$
and study only the dependence on $m_K$, in which case we find
\begin{equation}
  {m_{K^*}-m_\rho\over m_K^2-m_\pi^2}\to
  {{\rm d}m_{K^*}\over{\rm d}m_K^2}= {\lambda_2\over B_0} -{1\over12\pi f^2}
  \left[\left(4g_1g_2+4\sqrt2\,g_2g_4-9\sqrt 2\,g_2^2A_0\right)m_K
  +{7\sqrt2\over2}\,g_2^2\,{M_0^2\over m_K}\right],
\end{equation}
where in this expression $m_K$ is the one-loop corrected kaon mass.
The divergence of the final term as $m_K\to0$ reflects the unphysical
behavior of the quenched theory in the chiral limit.

Finally, we may estimate the size of quenching errors by comparing the
nonanalytic corrections to $m_\rho$ in the quenched and unquenched
theories.  Defining the ratio $R_\rho=m_\rho^Q/m_\rho^\chi$ to be the
ratio of the masses in the two theories, we find
\begin{eqnarray}
  R_\rho&=&1+\mu^{-1}\left[2(g_1^2-g^2_2+2g_2g_4)I_1(m_\pi)
  +4g_2^2I_2(m_\pi,m_\pi)\right]\nonumber\\
  &\simeq&1-{1\over6\pi f^2\mu}\left[m_\pi^3+3M_0^2m_\pi
  -5A_0 m_\pi^3\right]\,,
\end{eqnarray}
for $g_1^2-g_2^2+2g_2g_4=1$.  For the physical meson masses $m_\pi=140\mev$ and $m_\rho=770\mev$, this amounts to about a 30\% error from quenching (which, of course, is dominated by the uncertain $M_0^2$ term).  On the other hand, for larger lattice masses such as $m_\pi\approx500\mev$ and $m_\rho\approx1\gev$, the quenching error is of order 100\%.  Clearly, our estimate of the error depends on the many undetermined parameters which enter, but it does suggest that one ought to be cautious when using $m_\rho$ as a reference quantity to set the scale of the lattice spacing.

\subsection{The baryon-meson mass ratio}

\begin{figure}
\epsfxsize=12cm
\hfil\epsfbox{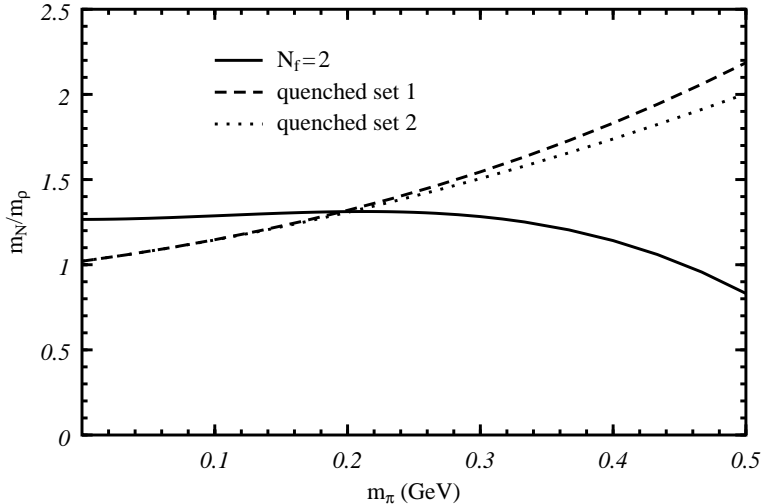}\hfill
\caption{The ratio $m_N/m_\rho$ as a function of $m_\pi$.  The vector meson parameters are as in Fig.~7, while the parameters for the baryon chiral Lagrangian are from Ref.~[9].}
\label{fig:massratio}
\end{figure}

The ratio $m_N/m_\rho$, which experimentally is 1.2, is used often
to gauge the accuracy of lattice simulations.  The chiral corrections
to $m_N$ have been computed by Jenkins~\cite{Jenkins:Bary} and by
Bernard, Kaiser and Meissner~\cite{Bernard:Bary1}, while the corresponding quenched corrections have been computed by Labrenz and Sharpe~\cite{Labrenz:QBary1,Labrenz:QBary2}.
Combining these with our calculation, we can 
examine the quenched chiral corrections to this ratio.  We find the expansion
\begin{equation}
  {m_{N}\over m_\rho} = {m_0\over \mu} + R_{1/2}\, m_\pi + R_1\, m_\pi^2 + 
  R_{3/2}\, m_\pi^3 + \ldots\,,
\end{equation}
with 
\begin{eqnarray}
  R_{1/2} & = & -\left[{3\over2}(D - 3F)^2 - 4g_2^2\right] 
	{M_0^2\over 8 \pi f^2 \mu} \\
  R_{1} & = & \left[2(b_D - 3b_F) - {\lambda_2\over B_0}\right]{1\over \mu} \\
  R_{3/2} & = & \left[(D-3F)(2D+\gamma) - c^2 + 4g_2(g_1+g_4)\right]{1\over12\pi f^2} - 
	{5 A_0\over 3 M_0^2}\, R_{1/2} \,,
\end{eqnarray}
where
$D$, $F$, $b_D$, $b_F$, $c$ and $\gamma$ are parameters of the 
baryon chiral Lagrangian~\cite{Labrenz:QBary1,Labrenz:QBary2,Jenkins:Bary,Bernard:Bary1}.
In  Fig.~\ref{fig:massratio} we compare the quenched and unquenched corrections, for a typical set of coefficients of order one.  Of course, there are far too many unknown parameters for this plot to be anything but illustrative of the possible size of the long distance effects.  What the reader should note is that once again there can easily be a large difference
between the two theories, and thus a large correction due to quenching.

\section{Discussion and Conclusions}

Let us close by commenting briefly on our results.  The most
important qualitative feature is the dependence of $m_\rho$ on
the {\em square root} of the quark mass $m_q$.
This is to be contrasted with the unquenched theory, where the
the leading dependence is linear and the leading nonanalytic dependence
goes as $m_q^{3/2}$.  
As noted earlier, while the coefficients of these terms depend on nonperturbative parameters, their presence is unambiguously predicted,
and they are unchanged by the inclusion of terms of higher order in the chiral expansion.  They dominate the extrapolation to small $m_q$, and must be accounted for in lattice simulations.

The derivation of our results has relied completely on chiral perturbation theory.  The careful reader might worry that
because quenched QCD is not a unitary theory, there is no guarantee
that QChPT accurately describes its low energy limit.  However, there is some independent evidence that QChPT is indeed the low energy theory of quenched QCD.  The most celebrated prediction of QChPT, the emergence of
``quenched'' logarithms of the form $M_0^2 \ln m_q$, was originally derived in another way, via strong-coupling perturbation theory~\cite{Morel:QLogs,Sharpe:QChPT1},
and the two approaches yield the same coefficient for these terms.
We thus expect that the use of QChPT to describe
quenched QCD at low energies is valid.

We should also mention one defect of this formulation of chiral perturbation theory for vector mesons, namely the absence of the decay $\rho \rightarrow \pi\pi$.  Clearly, this is an important contribution to the width
of the $\rho$.  However, its contribution to the real part of the
$\rho$ mass is rather small.  Moreover, a simple one loop calculation shows that the contribution is nonleading, of order
$m_\pi^4 \ln m_\pi$.  We thus expect that the inclusion of this effect would not alter our conclusions substantially.

Since our results depend on many unknown nonperturbative parameters, it is hard to draw firm quantitative lessons from them.  Nonetheless, we have found in general that the long distance quenching corrections to vector meson masses could be quite large, at the level of 100\%.  While it is equally possible that they are small, our analysis does nothing to increase one's confidence that quenching is a controlled error in lattice simulations of $m_\rho$ and $m_N/m_\rho$.  On the contrary, the necessity of simulations of these quantities with dynamical fermions is as pressing as ever.

\acknowledgements

This work was supported by the National Science Foundation
under Grant No.~PHY-9404057.  A.F.~acknowledges additional support from
the National Science Foundation for National Young
Investigator Award No.~PHY-9457916, the Department of
Energy for Outstanding Junior Investigator Award No.~DE-FG02-94ER40869,
and the Alfred P.~Sloan Foundation.

\bibliography{quenched}
\bibliographystyle{prsty}

\end{document}